\documentclass{sig-alternate}
\usepackage{times}
\usepackage[outercaption]{sidecap}
\usepackage{pgf}
\usepackage{tikz}
\usetikzlibrary{arrows,automata}
\usepackage[latin1]{inputenc}
\usepackage{amsmath}
\usepackage{graphicx}
\usepackage{amssymb}
\usepackage[tight]{subfigure}
\usepackage{pifont}
\usepackage{rotating}
\usepackage{booktabs}
\usepackage{multirow} 
\usepackage{xcolor}
\usepackage{url}
\usepackage{hyperref}

\definecolor{darkred}{rgb}{0.6,0.2,0.2}
\definecolor{darkblue}{rgb}{0.1,0.1,0.4}
\definecolor{mydarkblue}{rgb}{0,0.08,0.45}

{\bfseries}{\itshape}
{\bfseries}{\itshape}
{\bfseries}{\itshape}
{\bfseries}{\itshape}
{\bfseries}{\itshape}
\renewcommand{\P}{{\mathbf P}}
\renewcommand{\S}{{\mathbf S}}

\title{Unveiling Zeus}%
\subtitle{Automated Classification of Malware Samples}
\begin{document}
 \author{Abedelaziz Mohaisen\\
 \affaddr{Verisign Labs, VA, USA}\\
  \affaddr{amohaisen@verisign.com}
 \and 
Omar Alrawi\\
 \affaddr{ Verisign Inc, VA, USA}\\
  \affaddr{oalrawi@verisign.com}
 }


\maketitle

\begin{abstract}
Malware family classification is an age old problem that many Anti-Virus (AV) companies have tackled. There are two common techniques used for classification, signature based and behavior based. Signature based classification uses a common sequence of bytes that appears in the binary code to identify and detect a family of malware. Behavior based classification uses artifacts created by malware during execution for identification.  In this paper we report on a unique dataset we obtained from our operations and classified using several machine learning techniques using the behavior-based approach. Our main class of malware we are interested in classifying is the popular Zeus malware. For its classification we identify 65 features that are unique and robust for identifying malware families. We show that artifacts like file system, registry, and network features can be used to identify distinct malware families with high accuracy---in some cases as high as 95\%. 
\end{abstract}

\section{Introduction}\label{sec:introduction}

Malware family classification is age old problem that many industrial and academic efforts have tackled~\cite{tian2009automated,bailey2007automated,lee2006behavioral,park2010fast,tian2008function,rieck2008learning,zhao2010malicious,kinable2011malware,provos2007ghost}. There are two common techniques used for classification, signature based~\cite{rieck2011automatic,tian2008function,kinable2011malware} and behavior based~\cite{lee2006behavioral,park2010fast,rieck2008learning,zhao2010malicious}. Signature based classification is based on detecting a family of malware by using a common sequence of bytes that appear in the binary code. Behavior based classification is based on detecting a family of malware based on the artifacts the malware creates during execution. This paper will discuss a behavior based approach to classification of a single malware family, the Zeus banking Trojan~\cite{santos2011collective,ramilli2010multi,binsalleeh2010analysis}, using several machine learning algorithms. 

In this paper, the main malware family that we study is the Zeus Banking Trojan. The Zeus banking Trojan is a famous banking Trojan that is used by cyber criminals to run a botnet to steal money, credentials, and system resources from the infected victims. The Zeus source code was leaked in 2011 and since than there has been numerous variants that have surfaced~\cite{zeusurl}. Although the variants have new add-on features that are not found in the original Zeus banking Trojan, they all exhibit very similar behavior. Notable features in the new variants of Zeus include bitcoin mining, peer-to-peer command and control infrastructure, and added layers of encryption to the configuration file.

The Zeus malware infects the system by writing a copy of itself to the {\sf APPDATA} folder using a randomly generated file name. The stolen data is stored under the same directory, {\sf APPDATA}, encrypted~\cite{zbot}.  When Zeus sends the stolen data to the command and control server it deletes the local copy. 
Upon infection, Zeus injects into explorer.exe process and other running system processes to run out of. Zeus runs its main thread out of explorer.exe and communicates to the command and control server through explorer. New variants using peer-2-peer continue to run out of explorer.exe process but does not make any HTTP request. The Zeus banking Trojan hooks several important Windows APIs to intercept data being sent from the browser and to modify pages seen by the victim. For example Zeus is capable of adding fields in web forms to collect additional information from the victim when visiting banking site. 

After Zeus infects a system and establishes a connection with the command-and-control server, Zeus will download an updated version of the configuration file that tells the bot what sites to target. The configuration file use to be stored in the same {\sf APPDATA} directory as the rest of the files. The recent variants of Zeus have changed the storage area and storage method of the configuration to protected from being discovered. We have observed with the new variants of Zeus that the configuration file is encrypted and stored in the registry under a random key name. The configuration file is an important aspect of the Zeus banking Trojan and can reveal an abundant amount of information about the Zeus campaign. 

The configuration file can contain a list of backup command-and-control servers, link to an updated version of Zeus, list of targeted websites, and list of HTML and JavaScript to be injected in the targeted websites. The configuration file information is important because system administrators can block access to all domains used for backup, targeted entities can be notified about a particular campaign effecting their users, and security researchers can track the infrastructure used by the attackers.

The reason Zeus is an important piece of malware is because it is the most prevalent banking Trojan in the wild~\cite{zeusking}. Zeus accounts for most cyber crime targeting banks and small businesses, which calls for further investigation on identifying malware samples that belong to this family and exhibit new unique behaviors that we did not see before. To this end, this work is dedicated to unveiling Zeus; we identify a set of features in a large set of malware samples and use them automatically understand this important family.

To this end, the contribution of this paper is as follows. First, we report on our effort characterizing malware samples by automatically analyzing binary codes in isolated environments using one of our products, named {\sf automal}, and further discuss a small dataset obtained from this product. We identify a set of features that are representative to families of malware, including Zeus. Second, we use these features to automatically classify the different malware samples into families, using various machine learning algorithms, and report on both the efficiency and accuracy of the classification. 

The organization of the paper is as follows. In \textsection\ref{sec:prelim}, we discuss the preliminaries of this study. In \textsection\ref{sec:data}, we describe the dataset used in the study. In \textsection\ref{sec:experiments}, we report on the experiments, and highlight the accuracy and error when using different machine learning algorithms, some recommendations and observations. In \textsection\ref{sec:related} we review some of the related works followed by the future work in \textsection\ref{sec:conclusion}.

\section{Preliminaries}\label{sec:prelim}
In this study we used five classification algorithms to understand their capability in classifying different families of malware samples. In the following we review these algorithms.

\begin{itemize}\itemsep=-1mm
\item {\bf Support Vector Classification:} (also known as support vector machine; SVM) is a supervised deterministic binary classification algorithm that assigns a label to input determining which of two classes of the output it belongs to. Given a training set of samples (i.e., examples), each of which is marked as belonging to one of two categories, the algorithm builds a model that assigns each of the samples into one category or the other. The algorithm maps different samples as points in the vector space with a clear boundary between them, then the algorithm maps samples to the space, and associate them with other samples that are closer to them. The formal description of the algorithm can be found here~\cite{mlbook}. In the SVM algorithm, we use the L2 regularization (because of the particular settings of our dataset and the number of features) and L2 loss --- for more details see section~\ref{sec:experiments}.

\item {\bf Logistic Regression:}~\cite{mlbook} same as the SVM, logistic regression allows us to predict an outcome, such as label, from a set of variables. The goal of logistic regression is to correctly predict the category of outcome for individual cases using the best model. For that, a model is created that includes all predictor features that are useful in predicting the needed label. In our experiments, we use both L1 regularization  and L2 regularization --- In this study, we use both to identify the better one for the given dataset size and set of features. 

\item {\bf Classification Tree:}~\cite{mlbook} (also known as decision trees, or regression trees) is a model used for predicting the label by mapping observations about the sample to the conclusions about its target value. In the training part, the samples are used to create a model in which a boundary is created for each label based on the features. In the test part of the algorithm, the decision model created earlier is used for identifying which label is associated with the sample based on how it is fitted on the classification tree. 

\item {\bf K-Nearest Neighbor}~\cite{mlbook}  is a simple machine learning algorithm used for classifying samples based on the closest training samples to them in the training feature space. Because the algorithm is used for multiple classes, and our end goal is a binary classification of two classes, we limit the size of each cluster (class) discovered by the algorithm into the ideal size of each class in the training algorithm (in such case, we use equal size of samples in the training part; see details in the experiments). 

\end{itemize}
In the rest of this study we rely on an off-the-shelf implementation of the aforementioned algorithms. We use {\sf mlpy}~\cite{mlpy} (stands for Machine Learning Python), which is a python toolkit that implements several machine learning algorithms.

\section{Dataset}\label{sec:data}
A fundamental part in our contribution in this work is the dataset, the set of features included in the data, the way we obtain the dataset and the features, and the way we use for establishing a baseline for the ground truth by manual classification. In the following, we elaborate on the background of the dataset, the method used for extracting the raw features used for the classification, the method used for establishing a ground truth, and the baseline we use in our experiments, including the testing dataset.

\subsection{Background}
In this section we recall some of the background we mentioned in section~\ref{sec:introduction} about the Zeus dataset. For our data set we used Zeus Banking Trojan. As we mentioned earlier, the Zeus banking Trojan is a famous banking Trojan that is used by cyber criminals to run a botnet to steal money, credentials, and system resources from the infected victims. The Zeus malware infects the system by writing a copy of itself to the {\sf APPDATA} folder using a randomly generated file name. The stolen data is stored under the same directory encrypted in the {\sf APPDATA} directory.  In the following, we present the method used for extracting features and artifacts that are representative to the malware sample. 

\subsection{Raw and Vector Features Extraction}
The data is a set of 1,980 sample of the Zeus Banking Trojan. We ran these 1,980 samples through our automated malware analysis system {\sf auto-mal}. The system   {\sf auto-mal} is a virtual machine (VM) based system that is used to run samples of malware and capture behavior characteristics of that sample. The system enables us to set a run-time for each sample that long enough to capture enough artifacts about the sample; in our experiments we set the time into 1 minutes (upon several trials, we realized that 1 minute is enough for characterizing common samples). The  {\sf auto-mal} system uses several tools for capturing and characterizing networking traffic (IP address, port numbers, protocol types, and others --- for details see Table~\ref{tab:features}). Also, the  {\sf auto-mal} system uses tools, like sluethkit, which is used for file system and registry artifacts. In total, our system captures file system activities, registry activities, and network activities. The malware activity artifacts are logged to a MySQL database. We call those artifacts as raw features. 

From the raw features, we obtain a feature vector for each malware sample. Most of the features consist of counts and normalized data sizes that are used as features. For sizes, we consider the quartile counts (e.g., how many of the specified artifacts have a size that falls into the specified quartile of size generated by that malware sample; i.e., for the 1st, 2nd, 3rd, and 4th quartiles in relation with the file, for example, with the largest size). We have 65 features in total most of which are network features, as shown in Table~\ref{tab:features}.

\begin{table}[top]
\begin{center}
\caption{Features used in classifying malware samples.}\label{tab:features} 
\begin{tabular}{r|p{6cm}}
\toprule
Class & features \\ 
\midrule
{\bf File system} & created, modified, deleted, size (quartiles), unique extensions, count of files under common paths \\
\hline
{\bf Registry} & created keys, modified keys, deleted keys, count of keys with certain type \\
\hline 
{\bf Network} & see below for each sub-class \\
{\em IP and port} & unique dest IP, certain ports (18 ports)\\
{\em Connections} & TCP, UDP, RAW\\
{\em Request type} &  POST, GET, HEAD \\ 
{\em Response type} & response codes (200s through 500s)\\
{\em Size} & request (quartiles), reply (quartiles)\\
{\em DNS} & MX, NS, A records, PTR, SOA, CNAME \\
\bottomrule
\end{tabular}\vspace{-8mm}
\end{center}
\end{table}

\subsection{Sample Labeling}
The Zeus Banking Trojan samples have been identified by hand and collected over time by analysts---This process can be time-consuming. At average, a previously unseen malware sample (not necessarily Zeus) can take more than 10 hours to manually characterize by experts.  
The data sources are from various AV vendors that we have partnered with for sharing malware samples. The malware feed is delivered with no AV signatures associated with samples. We run Yara signatures on the malware feed to identify malware of interest that we can feed into our automated malware analysis system. 

We also have an AV scanner appliance that scans each sample going through our automated malware analysis system with 20 anti-virus scanners. This helps us identify if other vendors think a sample is Zeus, Zbot, or a different family of malware. Our automated malware analysis system has a memory forensics component that allows us to run Yara signatures on volatile memory to identify a specific family of malware based on strings in memory and byte sequences known for a specific malware family.  In the rest of this paper, we did not use memory features, and leave using them as a future work. 

\subsection{Baseline, Training, and Test Data}
The methods mentioned above are used in conjunction to identify and verify that each sample is of the same family and that we can use the behavior of these samples as features for the machine learning algorithms. The Zeus data set is split up into 2 different data sets one for learning and one for testing. The learning data set contains 1001 samples of Zeus and 1000 samples of other malware that is picked at random from our malware database collection. The testing set contains 979 samples of Zeus and 1000 samples of non-Zeus malware and that is not found in the learning data set. Notice that the number of samples we use in this study is only to illustrate the idea, and is way smaller than the total number of samples available to us. Classifying all samples is left as a future work.

\section{Experiments and Results}\label{sec:experiments}
In this section we discuss the experiments and the results of this study. Before going into further details, we outline the settings of the experiments and the evaluation metric. 

\subsection{Settings and Error Measures}
We ran the learning set, call it set A, through five different classification algorithms and tested the prediction on the testing set, call it set B. The family of linear classification had a cost of constraints violation set to 0.01. For the Class Tree classifier the minimum number of cases required to split a leaf is set to 5. For the KNN classifier the number of nearest neighbors are set to 980 (thus the number of the classes we have is $2$ for the $k$NN classifier). 

To evaluate and compare the different algorithms we use the false positive and false negative measures. The false positive error (false alarm; for the class label Zeus, for example) measures the number of samples marked as Zeus, while they are in reality not Zeus. On the other hand, the false negative error (for the Zeus family) measures the number of samples that are marked as non-Zeus, while they are in fact Zeus. Given that we are interested equally in both classes (Zeus and non-Zeus), and that the number of samples that are Zeus are not equal to the number of  non-Zeus samples, we generate the error measures for both of them as percents (normalized by the total number of samples in each class). 

\subsection{Results}

Using the settings above, we run these algorithms on our dataset, and computed the error measures (detailed in the previous section) normalized by the number of samples in each class. The results are shown in Table~\ref{tab:error1}. From those results, we observe the following (we verify some of those in the second experiment). First of all, we notice that the L1 regularization option when used with the logistic regression, is best suited for our dataset and the number of features we have to give the best results represented by the lowest error margin for both the Zeus and non-Zeus samples.  

Second of all, we notice that the support vector classifier provides the best results among all for the combined false positive and false negative measures, by identifying about 95\% of the Zeus samples correctly and missing only around 5\% of the samples at average, and by adding another 5\% misclassified samples to the final Zeus results. For the class of interest, the Zeus malware samples, while the decision trees algorithm provides the best (overall) false positive results, it provides a very high false negative (identifying samples as non-Zeus, while they are in reality Zeus samples) thus limiting its usefulness to the main purpose of the paper---correctly and accurately identifying Zeus samples. 

Finally, and while the false positive of the logistic regression (with L2 for regularization) is very large (around 27\% for both classes; perhaps because the number of the samples we have in relation with the number of the features is limited), we notice that the false negative provided by the algorithm is among the lowest in the study (about 2.5\%, yet higher than the L1 regularization case), which sheds light on its potential for identifying the class of interest, and perhaps limit its drawback by combining it with other algorithms that perform well for that class and that error measure. 
\begin{table}[top]
\begin{center}
\caption{False positive and false negative when running the different classification algorithms.}\label{tab:error1} 
\begin{tabular}{r|c|c}
\toprule
Algorithm &$+/-$ (Zeus) & $+/-$ (Non-Zeus)\\
\midrule
SVM & 6.84\%/4.29\% & 6.70\%/4.20\% \\
Logistic Reg. (L1) & 11.03\%/1.43\% & 10.81\%/1.40\% \\
Logistic Reg. (L2) & 27.06\%/2.55\% & 26.52\%/2.50\% \\
Decision Trees & 4.70\%/22.98\% & 22.52\%/4.60\% \\
KNN& 10.21\%/10.93\% & 10.71\%/10.01\% \\
\bottomrule
\end{tabular}\vspace{-5mm}
\end{center}
\end{table}

One challenging problem when using machine learning techniques for classifying data, particularly when using supervised learning techniques, is the choice of the training set. Starting with a well classified dataset (with respect to the classification features) may give nice results, or even starting with a poorly classified dataset might by misleading by showing superiority of an algorithm over another due to that fact, while in reality altering the initial learning dataset may greatly alter the findings.  To understand how our results are robust to the initial training set, we repeat the experiment by flipping the test and training datasets. We flipped the sets and made the B dataset as a learning set and the A dataset as a testing set, and ran the five different classification algorithms again to get the results shown in Table~\ref{tab:error2}.

In this experiment, we confirm the following (among the initial observations we made). First, we notice that the L2 regularization still performs worse than the L1 regularization with the logistic regression, establishing that the L1 regularization on the logistic regression is well suited for our dataset for the reasons listed earlier. Second, we observe that the (combined) performance of the SVM is still the best among all algorithms we used, which is perhaps due to the nature of our dataset: SVM performs well when the dataset consists of two clear classes, and that happens to be the case of our dataset. Third, unlike in the previous experiment, where the performance of the decision trees indicates a limited benefits of it, the false negative (of the Zeus class) when using the decision tree is greatly less (about 10\%) than in the previous experiment. Furthermore, the false positive is still same as in the previous experiment. This final observation shows how critical it is to start with a representative training set for the given classification algorithm. 

Fourth, we observe the symmetry in the false positive and false negative among the two classes and the different algorithms, which happens as a result of the equal size of samples that belong to each class in the testing dataset. Finally, we notice a limited difference in the results of the $k$NN classification algorithm, which is among the easiest to implement and run, indicating that, even when starting from a biased (or less representative training set), the $k$NN algorithm still provides reasonable results. 

\begin{table}[top]
\begin{center}
\caption{False positive and false negative when running the different classification algorithms.}\label{tab:error2} 
\begin{tabular}{r|c|c}
\toprule
Algorithm &$+/-$ (Zeus) & $+/-$ (Non-Zeus)\\
\midrule
SVM 				& 8.39\%/8.39\% & 8.39\%/8.39\% \\
Logistic Reg. (L1) & 7.29\%/8.29\% & 8.29\%/7.29\% \\
Logistic Reg. (L2) & 9.69\%/11.00\% & 11.00\%/9.69\% \\
Decision Trees 	& 4.90\%/12.79\% & 12.79\%/4.90\% \\
KNN				& 12.29\%/12.29\% & 12.29\%/12.29\% \\
\bottomrule
\end{tabular}\vspace{-5mm}
\end{center}
\end{table}

\section{Related Work}\label{sec:related}

There has been plenty of work in the recent literature on the use of machine learning algorithms for classifying malware samples~\cite{tian2009automated,bailey2007automated,rieck2011automatic,park2010fast,tian2008function,rieck2008learning,kinable2011malware,ramilli2010multi,provos2007ghost}. These works are broadly classified into two categories: signature based and behavior based techniques. Our work belong to the second category of these works, where we used several behavior characteristics as features to classify the Zeus malware sample. Related to work are the works in~\cite{rieck2011automatic,lee2006behavioral,park2010fast,rieck2008learning,zhao2010malicious}. In~\cite{park2010fast}, the authors use behavior graphs matching to identify and classify families of malware samples. In~\cite{rieck2008learning}, the authors follow a similar line of thoughts like ours for extracting features, and use SVM for classifying samples. Their ground truth relies on anti-virus reported classification which is mostly signature-based, and they do not include manual classification like in our case. Furthermore, the algorithms used for classification include only the SVM, which we tried along with other algorithms. Their dataset does not include any Zeus samples, and thus does not characterize this important malware family. The same work is extended in~\cite{rieck2011automatic}. Our work is different from the prior literature in two aspects. First, we limit our attention to understanding and classifying the Zeus malware sample, which is, to the best of our knowledge is not classified before. Second, to that end, our problem is limited in nature; we only use techniques that are designed for 2-classes classification problems, thus our error rates are smaller than those reported in the literature for multi-class classification.

\section{Conclusion and Future Work}\label{sec:conclusion}
In this paper, we have presented preliminary results on classifying Zeus, a popular malware family, using different machine learning algorithms. Much of the work is to be seen in the near future.
We are currently collecting a larger set of Zeus malware to be able to run the algorithms on a larger set. We are proactivily identifying Zeus banking Trojans and creating profiles for each new sample that comes in so we have a larger data set. We are also combing through our historical data to pull out samples that we might have missed and variants that might exhibited similar behavior but not exactly the same. We would like to run clustering algorithms on our entire data base of malware to break the samples into clusters then apply the memory signatures discussed earlier to label each cluster and identify sub-families within each of the classes. Finally, in the future we would like to combine different classification algorithms with different weights to improve the classifications results.

\end{document}